# Physical and Software Based Fault Injection Attacks Against TEEs in Mobile Devices: A Systemisation of Knowledge

**Joy, Aaron[1]; Soh, Ben[2]; Zhang, Zhi[3]; Parameswaran, Sri[4]; Jayasinghe, Darshana[5]**

[1]Tectilium, Australia. aj@tectilium.com

[2]La Trobe University, Australia. b.soh@latrobe.edu.au

[3]The University of Western Australia, Australia. zhi.zhang@uwa.edu.au

[4]The University of Sydney, Australia. sri.parameswaran@sydney.edu.au

[5]The University of Sydney, Australia. darshana.jayasinghe@sydney.edu.au

**Abstract**

Trusted Execution Environments (TEEs) are critical components of modern secure computing, providing isolated zones in processors to safeguard sensitive data and execute secure operations. Despite their importance, TEEs are increasingly vulnerable to fault injection (FI) attacks, including both physical methods, such as Electromagnetic Fault Injection (EMFI), and software-based techniques. This survey examines these FI methodologies, exploring their ability to disrupt TEE operations and expose vulnerabilities in devices ranging from smartphones and IoT systems to cloud platforms.

The study highlights the evolution and effectiveness of non-invasive techniques, such as EMFI, which induce faults through electromagnetic disturbances without physical modifications to hardware, making them harder to detect and mitigate. Real-world case studies illustrate the significant risks posed by these attacks, including unauthorised access, privilege escalation, and data corruption. In addition, the survey identifies gaps in existing TEE security architectures and emphasises the need for enhanced countermeasures, such as dynamic anomaly detection and updated threat models.

The findings underline the importance of interdisciplinary collaboration to address these vulnerabilities, involving researchers, manufacturers, and policymakers. This survey provides actionable insights and recommendations to guide the development of more robust TEE architectures in mobile devices, fortify FI resilience, and shape global security standards. By advancing TEE security, this research aims to protect critical digital infrastructure and maintain trust in secure computing systems worldwide.

## 1. Introduction

Trusted Execution Environments (TEEs) have become foundational in ensuring the security and privacy of sensitive operations on modern computing platforms. Designed as isolated secure zones in processors, TEEs provide a controlled execution environment for critical applications, safeguarding cryptographic operations, secure boot processes, and sensitive data from potentially compromised main operating systems. TEEs are integral to mobile devices,

IoT applications, and cloud platforms, underlining their importance in a connected and digitalised world.

However, the robust security guarantees of TEEs are increasingly challenged by advancements in fault injection (FI) techniques. FI attacks exploit vulnerabilities by introducing faults - either physically or through software manipulation - into a system to disrupt its expected operation. These faults can cause transient errors that bypass security mechanisms, resulting in privilege escalation, data leaks, or unauthorised code execution. Physical FI methods, such as Electromagnetic Fault Injection (EMFI) and Voltage Fault Injection (VFI), target hardware components to induce glitches, while software-based FI leverages vulnerabilities in application code or execution paths.

This survey aims to provide a comprehensive analysis of FI techniques used against TEEs, categorising them into physical and software-based approaches. It examines the methodologies, tools, and technologies behind FI attacks and explores their implications on TEE integrity. Furthermore, real-world case studies are analysed to highlight the effectiveness of these methods and the challenges in designing countermeasures.

The growing affordability and accessibility of FI tools further exacerbate the threat, making it crucial to understand the vulnerabilities in TEE architectures. This survey seeks to bridge the gap between existing FI methodologies and the development of robust defences, offering insights for researchers and practitioners aiming to fortify TEE security in the face of evolving attack vectors.

This survey analysis contributes to the following research outcomes:

- It provides a detailed analysis of FI techniques targeting TEEs, categorising them into physical and software-based methods. It highlights their effectiveness, methodologies, and tools, offering a structured overview for researchers and practitioners.
- It demonstrates the growing threat posed by non-invasive FI methods, such as EMFI, which are harder to detect and mitigate compared to traditional invasive techniques.
- Real-world examples are used to illustrate the practical risks of FI attacks on TEEs, detailing scenarios, such as privilege escalation, unauthorised access, and data corruption.
- It identifies specific vulnerabilities in current TEE implementations and gaps in existing security architectures, underlining areas that require further research and innovation.
- Actionable recommendations are provided to guide the development of more robust TEE architectures, incorporating dynamic anomaly detection, hardened designs, and updated threat models.
- Insights are offered for policymakers and standards organisations to establish guidelines for FI testing and to enhance the resilience of TEE-enabled devices.

## 2 Background

### 2.1 Fault Injection

Fault Injection (FI) is an array of hardware and software techniques that were developed for testing the dependability and security of modern computing systems. FI achieves this objective by targeting both the hardware and software of computing systems through the introduction of faults in a target system, causing the system to run outside the manufacturers recommended operating conditions [THI97]. While some FI techniques were developed in the 1970s, they still offer effective strategies for targeting modern security advancements in the form of TEEs, when compared to other techniques that are largely mitigated in modern secure architectures, such as Buffer Overflows and Man In the Middle attacks. Additionally, unlike the vulnerabilities discussed above, FI provides a means of identifying unique and often rare vulnerabilities prior to the release of a semiconductor in a cost effective way that enhances security at both a hardware and software level. The following provides a history of the hardware and software FI methods that have been used to target TEEs.

## 2.2 Hardware Fault Injection (HFI)

HFI was first used in the 1970s as a method for testing the durability of semiconductors with the aim of introducing abnormal and undesired behaviours [CCS99]. Initially, HFI techniques included Radiation- Based FI, Mechanical Stress Injection, and Temperature. Whilst these methods were focused on testing the reliability of semiconductors, their founding principles were later leveraged in the 90s as a means of bypassing security controls. Since the late 90s, the scope of HFI has expanded to include the use of EMFI, Clock Fault Injection (CFI), VFI and Optical Fault Injection (OFI).

### 2.2.1 Clock Fault Injection (CFI)

During the late 1990s a shift in the benefits of FI for testing the security of semiconductors took place, CFI grew in popularity, with notable early work including targeting smart cards [Moo+02]. Introducing CFI faults into a Device Under Test (DUT) relies on deliberately manipulating the clock signal of the DUT. CFI causes the timing of operations to fall out of sync often leading to timing errors, data corruption, instruction skipping, and repetition, etc. which can result in the bypass of security mechanisms if targeted correctly [CWE21; And01; Ago+10]. CFI has since been employed to exploit modern TEE architectures and other secure computing environments [TSS17]. One primary advantage of CFI is the precise control over the DUT clock, enabling faults (i.e. a slow down or speed up) to be injected at specific clock cycles to target a particular computational operation. This method is generally more cost-effective and poses lower risk compared to other glitching techniques discussed in this paper however uses a more invasive approach when compared to other HFI methods. Its invasiveness stems from the requirement to access a DUTs external clock for the manipulation of the a processors clock signal often requiring the addition or removal of components on the circuit board [SHO19].

### 2.2.2 Voltage Fault Injection (VFI)

Similar to CFI, VFI was initially leveraged in the late 90s to target the specific operating voltage ranges required for the reliable functioning of a semiconductor. VFI causes the semiconductors to function outside of the defined voltage parameters which can induce faults. One of the early uses of VFI in security testing was to target smart cards. Industry research in the 1990s and early 2000s focused on their implementation in Pay-TV boxes, targeting the extraction of the smart card cryptography keys by bypassing security checks performed by the card. One prominent example of this was [Aum+03] manipulating the voltage supply of a smart card, sending voltage spikes during cryptographic operations resulting in the signature verification process failing, this lead to a leak of information that was used to deduce parts of the private key. Unlike the other hardware FI methods discussed in this paper, VFI lacks the granularity of targeting a specific part of a semi-conductor, leading to system wide faults. VFI requires intrusive direct access to the DUTs power busses (similar to CFIs requirement for access to the external clock buss).

### 2.2.3 Optical Fault Injection (OFI)

OFI injection is a technique that uses a strong light source (such as a laser beam or a photo flash) to a DUT, inducing faults [vWM11]. OFI appeared at a similar time to EMFI, with one of the pioneering works being presented by [SA03] who decapsulated the PIC16F84 microprocessor. A high powered laser was then used to manipulate the behaviours of individual transistors in the PIC16F84 resulting in the bypass of security features in the secure microprocessor. In order for the successful injection of faults using OFI, the DUT needs to be decapsulated (i.e. using a strong etchant such as an acid), exposing its internal structures to the light radiation emitted by a laser diode [BH22]. The laser used to induce faults in a DUT is often attached to a XYZ stage for controlling the position of the laser relative to the DUT, allowing for the location of faults to be plotted, narrowing the search area after fault categorisation [Vas+20]. Although OFI has not been prevalent as a method for targeting TEEs, it has played an important role in the history of FI techniques used for testing device security. OFI requires direct access to the structure of the microprocessor to induce a fault which can be quite invasive.

### 2.2.4 Electromagnetic Fault Injection (EMFI)

EMFI is achieved through the use of strong electromagnetic fields, used to induce current in transistors and logic gates within the targets wafer, producing faults when induced at specific times and locations [DLM19]. Commercially available tools such as the Chip Shouter [1], eShard[2], and Riscure EMFI Transient Probes[3] are often combined with XYZ stages to control the location of the probe relative to the chip, allowing for the automation of fault location discovery within a System on Chip (SoC). Of the four HFI methods discussed, EMFI is the least invasive means of altering the code flow of a semiconductor as it does not require physical contact with the target or the removal of components from the target circuit board.

## 2.3 Software Based Fault Injection (SFI)

Like HFI Methods discussed in 2.2 SFI was first conceptualised in the 1970s, to determine the fault tolerance of computer systems. Some prominent research in the late 1980s and early 1990s demonstrated the development of software FI resulted in tools including Ferrari, DOCTOR, and FTAPE [KKA95; SHO19; TI95]. One of the core advantages of SFI over HFI is its non-invasive nature, in that attacks leveraging SFI techniques do not require physical access to a DTU or its circuitry. Common SFI techniques that have persisted from early research into methods that are still used to target TEEs include Code Mutation, Data Corruption, and Resource Exhaustion.

---

[1] https://www.newae.com/products/nae-cw520
[2] https://eshard.com/fault-injection
[3] https://www.riscure.com/products/em-fi-transient-probe/

### 2.3.1 Code Mutation (CM)

Code Mutation (CM) is a SFI technique appearing (slightly earlier than other SFI methods) in the 1980s, aiming to provide a technique for increasing the testing effectiveness for software developers. CM involves the modification of source and binary code, leading to the introduction of faults. Examples of CM faults include arithmetic such as changing num = f x p to num = f - p or logical statements such as if \$a > \$b to if a == b. These methods were developed to test code robustness, with [DeM+88] providing early work assessing the use of CM in the form of Mortha a software testing environment for inducing CM Faults [KO91]. Like the other HFI and SFI methods discussed in this paper, the focus later shifted from being purely for reliability testing, to a security focus in the 1990s. This change is outlined in [AKS96] with the use of code mutation for providing a novel approach to the use of CM to detect vulnerabilities that other techniques didn't reveal at the time.

### 2.3.2 Data Corruption (DC)

Data Corruption (DC) is a SFI technique developed in the 1980s and 1990s, aiming to deliver a methodology that provided programmers with the ability to simulate the corruption of memory in a target system and measure the impacts [GS95]. One of the pioneering works for DC was Fault Injection based Automatic Testing (FIAT), which was an early fault tolerance validation platform. FIAT achieved software testing through the introduction of faults, by directly injecting faults into target software (altering memory and register values) to simulate common faults. The faults injected included transient and permanent faults, causing faults by inducing bit-flips, stuck-at-faults, invalid memory access errors (by altering program counters), invalid stack pointers and values in memory [Seg+95]. These faults covered a range of likely software faults occurring in production. DC was later broadened to take on a security focus in the late 1990s to specifically focus on its use for injecting faults into UNIX command-line utilities by providing random inputs, leading to the corruption of the software as problematic values were injected. By observing the resulting crashes and errors, researchers were able to use their DC technique for vulnerability discovery [MFS89].

### 2.3.3 Resource Exhaustion (RE)

The concept of Resource Exhaustion (RE) testing first appeared around 1970, with Denning [Den68] providing some of the seminal defensive works aiming to assess methods for system optimisation. This was achieved through the proposed "working set model" which aimed to reduce the impacts of resource exhaustion on system performance by allowing for dynamic management of pages. This greater management of pages through the prediction of when they are required by a system, reduced page faults ultimately increasing system optimisations, which made it less susceptible to resource exhaustion FI. This was later used as a security testing mechanism, with academic discussion surrounding the use of RE for security testing beginning in the 1990s, with [Xu+01] demonstrating the use of *selective exhaustive injection* as a method for targeting specific code segments related to device security. The research conducted by [Xu+01] using RE FI resulted in the exhaustion of the target devices ability to handle legitimate requests (with 1-2% of faults violating security functions) allowing for research's to gain unauthorised access to the \$do_authentucation()\$ function in the SSH server binary running on the target.

## 2.4 Trusted Execution Environments (TEE)

TEEs are isolated partitions within semiconductors that aim to protect a subset of a device's memory and CPU using the concept of a trusted enclave [SAB15]. Trusted enclaves are hardware based security modules designed to only run code that is cryptographically signed, and encrypted, reducing the risk of the code being altered or read by any other code running externally from the trusted environment [Mic23]. The idea of a TEE was first researched in the late 1990s and early 2000s, with ARM first publishing documentation about their TrustZone platform in 2005 [ARM04] and Intel's LaGrande having technical documentation released in 2006 (superseded by SGX in 2013) [Int06]. Despite their innate benefits to the security of many consumer, commercial and military applications, one of the challenges with TEEs is their lack of standardisation, with each of the TEEs discussed throughout this paper having vastly inconsistent and undefined implementations of core security concepts [Lee+20; SAB15]. Some of the major TEEs used throughout modern computing platforms include but are not limited to:

- ARM TrustZone
- Intel SGX
- OP-TEE
- AMD Secure Encrypted Environment (SEV)
- Qualcomm Secure Execution Environment (QSEE)

### 2.4.1 ARM TrustZone

ARM's implementation of a TEE, was first released in 2004 as TrustZone, aiming to provide a scalable security platform for integration with billions of ARM devices globally [AMD12]. Unlike the other TEEs discussed in detail, ARM TrustZone employs System-wide segmentation of system resources into separate 'secure' and 'non-secure worlds', effecting the security of the entire system, which is prominently implemented in the ARM Cortex-A and Cortex-M processors [ARM19]. A processor that has ARM TrustZone implemented can switch between the secure and non-secure states with sensitive tasks handled in the secure state and non-secure tasks handled by the non-secure state. Additionally, the implementation of the 'secure' and 'non-secure' states is applied separately from privilege, i.e. a user with root privilege does not hold any additional ability to manipulate code running in the processors secure state. The transition between states without compromising security is possible through secure registers that are not duplicated, allowing for information to be passed between states and upon switching states the ability to restore register values [Nga+16].

### 2.4.2 Intel Software Guard Extensions (SGX)

Intel SGX, introduced in 2013, is designed to enhance the security of Intel's architecture. SGX primarily focuses on ensuring the integrity and confidentiality of sensitive computations at the kernel and hypervisor levels, addressing key aspects of the Confidentiality, Integrity, and Availability (CIA) triad [CD16]. SGX provides developers a platform for executing their code within a secure enclave, providing separation and encryption of sensitive processors protecting secure memory [Int22]. It achieves this through software, by only allowing read and write protection from inside the enclave regardless of privilege, debug protection, isolation from function calls, alterations to code flow, register manipulation, or stack manipulation and only supports 32-bit and 64-bit modes for enclave code execution.

Furthermore, Hardware protections include enclave memory encryption with replay protection, memory encryption in transit, memory encryption changes on power cycle, with keys stored inside the CPU [CD16].

### 2.4.3 Open Portable-TEE

First developed by ST-Ericsson for the ARM architecture, Open Portable TEE (OP-TEE) was released as open source in 2014 [tru23]. OP-TEE assumes that the OS is non-secure, protecting trusted applications in a light weight and modular way Through the use of secure user space libraries, a secure privileged layer, TEE Client APIs, and LinuxTEE kernel driver to achieve code segregation [tru24].

### 2.4.4 AMD Secure Encrypted Environment (SEV)

In 2017 AMD released technical documentation [Kap17] pertaining to their Secure Encrypted Environment (SEV) that focus on addressing segmentation of Virtual Machines (VMs) from one another and the Hypervisor (running on an AMD platform). SEV achieves segregation through the implementation of an AMD Secure Processor (ASP) that creates a unique encryption key for each VM, providing memory encryption (AES-128) that is managed by the processor. Additionally, SEV provides Key Management (with access segmented from the hypervisor), implementing SEV-Secure Nested Paging (SNP) and secure boot which aims to provide users additional integrity assurances that provide measurable and verifiable system state metrics [AMD20].

### 2.4.5 Qualcomm Secure Execution Environment (QSEE)

Qualcomm first implemented QSEE TEE in 2011 using the ARM TrustZone TEE as a framework into their Snapdragon MSM8x60 and APQ8060 processors [Qua11]. Since 2011 Qualcomm has continued to develop the QSEE which has been implemented in most modern Android Smart Phones such as Pixel, LG, Xiaomi, Sony, HTC, OnePlus, and Samsung. While Qualcomm has also deployed QSEE in other industries such as IoT and Automotive [Mak19]. QSEE offers a Secure Enclave through the implementation of a Secure File System (SFS) that encrypts a devices file system for use by the TEE, Replay Protected Memory Blocks (RPMB), which is a secure partition on an eMMC or UFS module that prevents replay attacks, Secure Processing unit, implementing an independent boot-loader, boot chain, anti-replay protection and cryptographic management module. Additionally, QSEE enforces strict isolation between trusted and un-trusted applications through sand-boxing and metadata encryption, preventing unauthorised access from un-trusted applications [Cai19].

### 2.4.6 Other notable TEEs

In addition to the TEEs discussed in detail above, the following outlines other prominent TEEs and their typical use cases:

1. Apple Secure Enclave (ASE) - Used in the IOS platform for biometric and key management, specifically Touch ID/Face ID as well as encryption keys and other sensitive information [App24].

2. Microsoft Virtual Secure Mode (VSM) - virtualisation based security features in windows; credential and device guard, and the execution of security sensitive applications [Mic22].

3. Huawei TEE - Secure environment for the execution of trusted platforms used for mobile payments, DRM and secure communications [BWM20].

4. Trusty - Androids TEE designed to run on the same ARM hardware as TrustZone (using Trust- Zone's virtualisation technology) and Intel (Using Intel's Virtualisation Technology) using minimal OS kernel. Additionally, it leverages a Linux kernel driver for secure communication and a secure API for trusted applications and manages keys and authentication throughout the system [And24; Kuh+24].

5. Keystone Enclave - An open source TEE developed for RISC-V processors, enforcing hardware-enforced and software-defined memory isolation with a platform agnostic approach, requiring minor effort to implement it into a wide range of RISC-V platforms [Lee+20].

## 2.5 Related Work

### 2.5.1 A Survey on the (In)Security of Trusted Execution Environments

-[Mun+2]

Provides an in-depth analysis of the challenges facing the security of modern TEE architectures. The paper provides in-depth analysis and categorisation of the vulnerabilities discovered against TEEs and the design implementations in each that lead to them. Its relevance to this paper is its assessment of some of the Fault Injection Methodologies targeting TEES, including hardware-based attacks like BADFET and the increasing prevalence of EMFI, some of the emerging software fault injection at- tacks targeting Dynamic Voltage and Frequency Scaling (DFVS) (CLKSCREW and VoltJockey) and Rowhammer. These papers are surveyed through the use of a Literature Review structure, categorising the characteristics of attacks into software-based, architectural, side-channel and micro-architectural attacks, which is broken into targeted categories including kernel and system call attacks and imple- mentation bugs. The impact of each of these attacks is assessed based on their impact to the target TEEs security and how design flaws were exploited by attacks, and how counter measures can be ap- plied to mitigate their impact. Finally, the paper suggests the challenges in the area of TEE Security, proposing areas of future research.

### 2.5.2 Physical Fault Injection and Side-Channel Attacks on Mobile Devices

- [She+22]

Focuses on the analysis of Fault Injection Attacks (FIAs) and Side-Channel Attacks (SCAs) against Mobile Devices, with a focus on Trusted Execution Environments. The paper reviews various implementations of TEEs with a heavy focus on TrustZone. This assessment outlines the TEEs features and evolution overtime, categorises their vulnerabilities and attacks (architectural, side-channel and micro-architectural) as well as countermeasures that mitigate the effectiveness the discussed attacks as well as challenges and recommendations. Key findings from this paper include the Increased use of Side-Channel Attacks to target

Intel SGX and Attacks against ARM virtualised systems and timing attacks against Android smartphones. Furthermore, the paper discusses Branch Predictor and Cache-Timing Attacks (Prime Probe and Evict Time attack) and Memory Exposure and Boomerang attacks (existing in the communication between secure and non-secure realms). Finally, the paper outlines open challenges that pose new research direction, focusing specifically on the protection of shared secrets between realms, continuing side-channel prevalence and potential solutions.

### 2.5.3 How Practical are Fault Injection Attacks

- [BH22]

Outlines various fault injection attacks (hardware and software), considering their viability in targeting Cryptographic Algorithms, Integrated Circuits (ICs), Microcontrollers, TEEs, Neural Net- work Implementations, Standard Computing and Embedded Systems and Memory Components. The survey considers factors including cost, technical complexity, the impact on security and the real-world feasibility of the attack. These factors assist in the papers determination of whether highly specialised labs are required or whether the techniques and attacks are feasible for individuals. The papers methodology focused on an in depth review of hardware fault injection attacks (CFI, EMFI, OFI, VFI) and software attacks like Rowhammer and the required hardware to target various architectures including AMD, ARM and Intel as well as an overview of their use to attack these architectures and corresponding countermeasures. The paper concludes that with low-cost equipment (roughly $1000) many of the discussed fault injection methods can be repeated, with [BH22] suggesting that VFI can be repeated with hardware such as the Teensy 4.0 board to target AMD and Intel security enclave and CFI can be relatively easily repeated. However, with other methods like OFI and EMFI greater level of expertise is require and financial investment is often required. Whilst this paper does cover TEEs, it covers a much broader range of targets. Additionally, it places more of an emphasis on the Hardware Fault Injection methods, with this paper focusing equally on the SFI and HFI techniques.

### 2.6 Non-Invasive Fault Injection Attacks

Assesses non-invasive fault injection methodologies targeting microelectronic systems, including IoT devices, cryptographic circuits, micro controllers and microprocessors, cloud computing infrastructure, TEEs and Hardware Accelerators. The paper categorises various non-invasive fault injection methodologies, comparing the impact that they have on the various target systems outlined above and evaluates existing assessment frameworks and mitigation strategies grouping them based on implementation levels and effectiveness. The papers assessment covers voltage and clock glitching attacks against TEEs (CLKSCREW, Plundervolt), side channel attacks, SFI covering DVFS and Remote Timing Fault attacks, exploiting race conditions, inducting faults in TEEs that leak sensitive data. Ultimately, the paper found that non-invasive fault injection attacks pose a severe risk to the target devices assessed due to their increasing affordability and ability to exploit a target without physically tampering with it, emphasising the need for more research in this space based on the gaps they identified. They also deliver a taxonomy that categorises the discussed noninvasive fault injection techniques and the need to build on their research to develop robust mitigation strategies and devices assessment frameworks.

# 3 Attacks Against TEEs

## 3.1 Hardware Fault Injection Targeting Trusted Execution Environments

Most of the recent literature assessing hardware fault injection to target TEEs began in 2017, with contribution made by both industry and academic researchers. Traditionally for a successful fault injection attack, a 'trigger' is required that signals the optimal time to perform a glitch, as seen in many of the researched papers outlined below. For instance, when targeting a secure boot function regardless of the HFI method used, a classification of the correct time to inject the fault is often critical for a successful fault injection.

### 3.1.1 Trigger Mechanisms

Side Channel:

A significant challenge for vendors in the development of safe and secure hardware devices is the reduction in the leakage of sensitive information from a given semi-conductor [Spr+18]. This is problematic, because an attacker may be able to infer information like the time a cryptographic function is running and use this to trigger a fault injection attack. Much of the research conducted that target TEEs using HFI, rely on the leakage of information, with the following examples:

1. Electromagnetic Radiation Emanations (ERE) – Unintended electromagnetic signals dissipated from a semi-conductor that reveal sensitive information, such as the timing of process execution within a TEE, like cryptographic functions [Lem+19]. One of the commercially available tools offered for the detection of EREs is the Riscure Transceiver which is used for triggering fault injection in the Riscure EMFI ecosystem [Ris24].

2. Power Analysis (PA) – Unintentional leaks of a semiconductors current and/or voltage during key operations like cryptographic functions (such as secure boot) with mainstream methods including Simple Power Analysis (SPA) and Differential Power Analysis (DPA) [KJJ99; CMW14].

3. Timing Analysis (TA) - By analysing the timing of code execution, it is possible that an adversary can trigger a certain event, wait a precise time and then trigger a FI for the purpose of bypassing particular security functions [LN18].

4. Hardware Interfaces (UART, JTAG, SWD, etc) - Inadvertent access to debug interfaces on target devices that provide access to log information that can be leveraged for timing FI attacks [FAF06].

Whilst there are many more methods used in industry for side channel analysis, the four outlined above are the most used in the HFI research discussed in this paper.

### 3.1.2 Targeted Components

Each TEE has its own set of operations that it runs within its secure enclave and in the case of the TEEs discussed in this research paper the components that are most targeted:

1. Secure Boot Functions – functions that perform cryptographic hashes on a boot loader to ensure its integrity and authenticity, aiming to guarantee that only signed firmware can be booted. This aims to prevent malicious software from being loaded on a device [Haq+20].

2. Cryptographic Functions – code that perform cryptographic hashes and other integrity checks running within the TEE aiming to prevent code running in untrusted processor cores from accessing data designated for the secure enclave [Bar+12].

3. Restricted shell access – Security checks implemented to prevent access to debug interfaces such as JTAG, SWD, UART, etc. As well as shell access provided by bootloaders such as UBoot that all provide varying levels of command line access to a target IC [CH17].

### 3.1.3 Case Studies

Some of the notable papers that evaluate the use of HFI against TEEs that are assessed in this paper include:

1. [Lim20] focused on the use of VFI to target the ARM TrustZone TEE running on a Nuvoton M2351 (Coretex-M23). The trigger for the attack is through identifying voltage spikes detected during cryptographic operations using an oscilloscope, this is used to trigger the VFI glitches against the Coretex M23. The paper does not allude to the success rate of the attack. The indicated cost of the DIY glitching board used in the attack is 2 Dollars.

2. [CH17] researched the impacts of a low-cost DIY EMFI setup using a EMFI probe attached to a 3D printer gantry. The attack is triggered by timing analysis (by determining that 4.62 seconds after a power cycle) targets the Broadcom BCM11123 in the a Cisco 8861 IP phone leading to the bypass of the phones secure boot checks at a distance of 3mm from the semi-conductor. The attack was successful in triggering a UBoot shell 72% of the time (72 of the 100 times) that the BCM11123 was glitched. The benefit of this research is its focus on the use of DIY, non-commercial, cost effective, and high performance EMFI for security testing of a target devices.

3. [Fan+23] targets an ARM Cortex A53 Smartphone development board, using EMFI. The attack is triggered through the use of a Software Defined Radio (SDR), placed in close proximity to the target semi-conductor allowing for real-time detection of ERE side channels to synchronise glitch timing during the Linux Kernel Authentication stage of Android secure boot. Ultimately, this resulted in Bypassing the authentication mechanism of Android Secure-Boot with a success rate of 0.55% (83 of 15000 injections successful or one successful bypass every 15 minutes). The benefit of this approach is that the technique can be used in either a black or grey box context making the research applicable to untested environments.

4. [Rae21] leveraged EMFI to target the QSEE TEE running on the Qualcomm IPQ40xx. The researchers were able to use the devices UBoot interface to run a characterisation script, that outputted an expected value over UART, with an unexpected result, indicating a fault. The researchers indicate that 5% of faults (with one experiment every 20 seconds) were successful in bypassing range checks.

### 3.2 Software Fault Injection Targeting Trusted Execution Environments

When compared to HFI methods for inducing faults in TEEs, research detailing SFI methods hasn't been explored to the same extent. Traditionally, SFI methodologies rely solely on the use of software-based attacks to target a semiconductor, piece of code, etc., whereas SFI used to target TEEs often demonstrates the use of a hybrid methodology combining both HFI and SFI to target a chip (with roughly half of the research covering SFI research against TEEs in this paper leveraging this approach). This hybrid approach for targeting TEEs relies on the use of a software vulnerability that provides access to a DUT's system functions that are used for controlling clock or voltage. By abusing access gained to these system functions, researchers are able to leverage them to enable a HFI attack (such as VFI or CFI) to target the secure enclave running within a TEE.

The software vulnerability can target either a function within the processor or an external auxiliary component on the circuit board, surrounding the processor to achieve this purpose. In addition to the hybrid approach, attackers have leveraged Cache Based FI attacks (purposefully altering cache components that exist in unsecured memory, that are accessed by applications in a TEE) and row hammer based techniques (to flip bits in the target TEE). Most SFI techniques targeting TEEs aim to manipulate a shared domain between the un-trusted and trusted cores to impact the trusted core within a processor.

### 3.2.1 Model Specific Registers (MSRs)

One of the common techniques leveraged by researchers to target the Intel SGX TEE was the manipulation of the platforms MSRs, which are specific registers used by the CPU to control a variety of its components that contribute to the processors efficiency and performance [Int14].

Common attacks exploiting MSRs in SGX focus on registers that are responsible for controlling the processors voltage and clock parameters which were accessible from the un-trusted core, and their manipulation could impact the secure enclave within SGX [Kog+22]. For Instance, an attacker in the un-trusted core that has root privileges can issue commands to a specific MSR that allows for the voltage and clock of the trusted core to be controlled, compromising confidentiality and integrity components of the SGX TEE.

### 3.2.2 Dynamic Voltage and Frequency Scaling (DVFS)

Similar to the Intel SGXs MSR vulnerabilities, ARM TrustZone has seen vulnerabilities with their DVFS functionality, which are used for controlling the clock frequency and voltage of a processor (designed to increase battery and processing performance) [GKS22]. Like the MSR vulnerability with SGX, commands run in the un-secure world of a processor running ARM TrustZone can affect the voltage and/or clock of the trusted core inducing HFI like faults using software vulnerabilities.

### 3.2.3 Shared Caches

Unlike the DVFS and MSR targets, cached based FI targets the shared nature of cache files in some TEE implementations. Manipulation of the shared cache from the non-secure state may result in the secure-state caching sensitive data that is observable to the non-secure state [OST05]. A common example of this form of FI is the flushing of certain cache lines that force a secure process to fetch data from memory, disrupting its operation or revealing how the process is functioning. The impact of this form of vulnerability is the breakdown of the TEEs confidentiality element through its impact to the TEEs isolation between worlds [Ge+18].

### 3.2.4 Trigger Mechanisms

Unlike attacks that focus on the use of hardware to induce faults in TEEs, software based FI Attacks, whether purely software or a hybrid software and hardware attack, provide more

control over the trigger of the fault. SFI triggers used for inducing a faults into TEEs include:

1. Timing-Based Triggers (TBT): Running target code inside a secure enclave, then altering the system to induce faults, that lead to the leaking of sensitive information from a TEE, i.e. altering the system whilst a cryptographic function is running [JSM07].

2. Event-Based Trigger (EBT): Inducing an event that leads to a fault to alter the control of the software and/or the interception and modification of system calls to cause faults an important points in code flow [Gir05]. One of the prevalent examples of an event-based trigger is Cache- Based FI, invalidating target cache lines with stale data, leading to the recovery of sensitive information from a targeted process, relying on specific conditions to induce a fault.

3. Environment Based Manipulation Trigger (EBMT): This trigger encompasses the use of external processes like MSRs and DVFS or auxiliary components that are used to control components that impact the operating environment of a targeted TEE. The modification of these external components can lead to fluctuations in voltage and clock signals causing an internal fault within a target TEE [Moo+02].

### 3.2.5 Targeted Components

1. Cryptographic functions – Researchers have used trigger timing methods such as TBT to target cryptographic functions as they are running on a TEE, causing them to leak secrets such as keys [YSW18].

2. Device functionality – Access gained to the software power control components on a circuit board, used to deliver destructive voltages, destroying processors running TEEs, acting as a denial-of-service attack [MG17].

3. Code Execution – leveraging access vulnerabilities in the implementation between the secure and non-secure cores to induce faults that manipulate target functions, allowing for unauthorised access [Cot+12].

### 3.2.6 Case Studies

The following case studies demonstrate SFI attacks against Trusted Execution Environments.

1. [CO23] demonstrates the vulnerabilities created by unsecured auxiliary components on the 'moth- erboards' surrounding a processor, using access to a BCM IC to under-volt an Intel processor running SGX. Vulnerabilities in the Baseboard Management Controller (BCM) which were exploited to gain access to the PMBUS which is responsible for the Voltage Regulator Module (VRM) which delivers core voltages to the Intel CPU running SGX. By manipulating voltages, researchers were able to under-volt the SGX core, triggering faults. This attack is controlled by the attacker, iteratively lowering core voltages as target processes are executed in the SGX core. By altering the voltage delivered to the secure SGX core, attackers were able to induce Data Corruption faults resulting in, (of the 253 tests conducted over a period of 535 mins), 194 faults, of which 66 leading to the recovery of RSA private keys (an overall success rate of 26%).

2. [Zha+24] outline the use of Cache-Based Fault injection (CBFI) vulnerabilities in the AMD SEV TEE and in this instance, there is no specific trigger required to perform the glitch against a target Virtual Machine (VM) running with AMD SEV. The CBFI attack caused the resource exhaustion of the AMD SEV's Cache function, allowing attackers to gain root access to target VMs by recovering the private keys used for authenticated SSH sessions and then elevating privileges. The success of recovering the private key was recorded 90%, the take-over of the authenticated SSH sessions was 100% and privilege escalation resulting from targeting the setuid binary was successful 99% of the time. Event-Based Trigger (EBT)

3. [TSS17] leverages the use of a hybrid SFI and HFI attack, targeting the DVFS power management process controlling the target ARM CPU's voltage, leading to control over the clock cycles sent to the ARM processor running in the target Nexus 6. The Code Mutation faults induced by CLKSCREW are caused by controlling the precise timing of clock and voltage signals at key moments during the execution of cryptographic function leading to instruction skilling, incorrect instruction execution and incorrect data passing. Despite similarities to other attacks outlined in this paper, in that CLKSCREW used a Hybrid HFI and SFI approach, the requirement for the timing of the attack to specifically run during cryptographic functions, makes its categorisation fall under a Timing, rather than Environment Based Trigger. The attacks included the inference of an AES key, with 60% of faults precise enough to cause a fault and the second attack being more generalised fault injection, leading to 1553 of 6000 attempts (roughly 20%) resulting in successful faults.

4. [Jan+17] demonstrated the use of the SGX-Bomb attack, leveraging the row hammer technique to target the Intel SGX TEE, causing a denial of service to the targeted Intel Core I7-6700K. The attack identifies the same DRAM bank that conflicts with one another, leveraging timing analysis and side channels to find these target rows in the enclave, focusing on the differing access times for rows in the same DRAM bank when compared to other DRAM banks. Once the attacker finds conflicting rows, row hammer is used to induce data corruption faults leading to bit flips in adjacent rows, accessing conflicting rows while bypassing the cache (leveraging cache flush instruction) to enhance the likelihood of bit flips. The induced bit flips cause integrity check failures in the Memory Encryption Engine (MME) as the SGX evolves memory is accessed causing the drop-and lock policy to engage, leading to a system wide Denial-of-Service Attack. Unfortunately, the success rate of this attack is not explicitly discussed.

## 4. Discussion

Trusted Execution Environments (TEEs) represent a cornerstone of modern secure computing, providing a dedicated, isolated space in processors to execute sensitive tasks and protect critical data. However, as this survey highlights, TEEs are increasingly vulnerable to sophisticated fault injection (FI) techniques, including physical and software-based attacks. These vulnerabilities pose significant risks to the confidentiality, integrity, and availability of sensitive operations, particularly in devices that underpin modern digital infrastructure, such as smartphones, IoT devices, and cloud platforms.

One of the key findings of this analysis is the growing accessibility and effectiveness of Electromagnetic Fault Injection (EMFI) and Voltage Fault Injection (VFI). These non-invasive techniques allow attackers to induce transient faults without requiring physical alterations to the target hardware, making them harder to detect and mitigate. For example, EMFI exploits electromagnetic pulses to disrupt the normal operation of electronic circuits, leading to corrupted memory contents or unauthorised privilege escalation. Despite the advancements in TEE security, these methods expose inherent weaknesses in the hardware trust model, highlighting the need for improved resilience against physical FI attacks.

Software-based FI techniques also present a unique challenge, leveraging vulnerabilities in the execution paths or algorithms of TEE applications. These attacks can bypass security measures, such as secure boot processes or cryptographic protections, exploiting logical errors in the software. The dual focus on physical and software-based methods underlines the complexity of defending TEEs, as vulnerabilities can emerge from both layers, requiring holistic security solutions that integrate hardware and software countermeasures.

The implications of these findings extend beyond individual devices to broader cybersecurity and national security concerns. The increased affordability of FI tools lowers the barrier to entry for potential attackers, making it imperative to prioritise the development of robust mitigation strategies. This survey identifies several areas for improvement, including enhanced fault detection mechanisms, hardened TEE architectures, and comprehensive testing frameworks. In addition, the integration of advanced monitoring systems capable of identifying anomalous behaviours caused by FI attacks is essential for real-time threat mitigation.

Moreover, the ethical considerations surrounding FI research are pivotal. While fault injection can reveal critical vulnerabilities, the dissemination of such findings must be carefully managed to prevent misuse by malicious actors. Collaborative efforts among researchers, manufacturers, and policymakers are crucial to ensuring that the insights from FI research translate into actionable improvements without inadvertently increasing risks.

## 5. Implications

The findings of this survey have profound implications for the future of hardware and software security, particularly in the context of Trusted Execution Environments (TEEs). TEEs play a critical role in safeguarding sensitive data and processes across a wide range of devices, including mobile phones, IoT systems, and cloud platforms. However, the vulnerabilities exposed by fault injection (FI) techniques, such as Electromagnetic Fault Injection (EMFI) and software-based FI, highlight the urgent need for enhanced security measures to protect these essential components.

The demonstrated effectiveness of non-invasive fault injection techniques, such as EMFI, underscores the need for manufacturers to adopt more robust design principles. Implementing fault-resilient architectures, adding electromagnetic shielding, and integrating real-time anomaly detection systems are necessary steps to mitigate the risks posed by these advanced attack methods. Manufacturers must also prioritise regular security testing using FI tools to identify and address vulnerabilities before devices are deployed.

This survey reinforces the importance of updating security standards to reflect the growing sophistication of FI techniques. Policymakers and standards bodies should work collaboratively to establish guidelines that mandate rigorous fault injection testing during the design and production stages. These guidelines should also include requirements for post-deployment monitoring and timely patching of vulnerabilities identified through FI research.

Ensuring compliance with these standards will enhance the overall security posture of TEE-enabled devices, reducing risks to users and critical infrastructures.

The insights from this survey highlight significant opportunities for advancing research in fault injection and TEE resilience. Developing more effective simulation frameworks, refining FI techniques, and exploring innovative countermeasures are essential to staying ahead of emerging threats. The interdisciplinary nature of FI research necessitates collaboration across fields, including hardware engineering, software development, and cybersecurity policy, to create holistic solutions.

The vulnerabilities identified in TEEs have far-reaching implications for national security and the protection of critical infrastructure. Devices secured by TEEs are increasingly used in defence, healthcare, finance, and other vital sectors. The potential exploitation of these vulnerabilities by malicious actors could result in severe disruptions or data breaches. Addressing these risks through robust TEE security will enhance the resilience of national infrastructures against sophisticated cyber threats.

The ethical implications of FI research cannot be overlooked. While fault injection techniques are powerful tools for uncovering vulnerabilities, their misuse could pose significant risks. Responsible disclosure practices and collaboration with device manufacturers are essential to ensure that the findings from FI research lead to strengthened security rather than increased exposure to threats.

## 6. Recommendations

This survey identifies critical vulnerabilities in Trusted Execution Environments (TEEs) arising from fault injection (FI) techniques, including Electromagnetic Fault Injection (EMFI) and software-based methods. To address these risks and bolster TEE security, the following recommendations are proposed:

- Enhance TEE Design and Development
    - TEE designs should integrate fault-resilient architectures, including error detection and correction mechanisms, to mitigate the impact of transient faults induced by FI.
    - Enhance the isolation capabilities of TEEs to prevent faults from propagating across secure and non-secure environments, thereby safeguarding sensitive data.
    - Physical countermeasures, such as EMI shielding, should be employed to minimise the susceptibility of TEEs to non-invasive fault injection techniques.
- Adopt Rigorous Testing Frameworks
    - Manufacturers and researchers should adopt advanced simulation frameworks to model potential fault injection scenarios during the design and testing phases.
    - Devices should undergo rigorous FI testing, including both invasive and non-invasive techniques, to identify vulnerabilities under diverse attack conditions.

- Develop Real-Time Detection and Response Systems
    - Integrate real-time monitoring systems capable of detecting anomalies caused by FI attacks, enabling devices to respond proactively to mitigate potential damage.
    - Implement adaptive countermeasures that can reconfigure security settings or isolate affected components in response to detected FI attempts.
- Update Threat Models and Security Standards
    - Existing threat models should be updated to account for the growing prevalence and sophistication of non-invasive FI techniques, such as EMFI.
    - Policymakers and standards organisations should develop benchmarks for evaluating and certifying the resilience of TEE-enabled devices against FI attacks.
- Foster Collaboration Across Stakeholders
    - Encourage collaboration between academic researchers, industry practitioners, and device manufacturers to share insights and develop comprehensive solutions.
    - Work with government agencies to establish regulatory frameworks that mandate FI resilience in TEE designs, ensuring alignment with national security priorities.
- Promote Ethical Research Practices
    - Researchers should follow responsible disclosure practices when identifying vulnerabilities, providing manufacturers with sufficient time to implement fixes before publicising findings.
    - Limit access to advanced FI tools to authorised researchers and organisations to prevent misuse by malicious actors.
- Advance Educational and Training Initiatives
    - Develop training programs for engineers and cybersecurity professionals focused on FI techniques and mitigation strategies.
    - Educate manufacturers and policymakers about the risks associated with FI attacks and the importance of adopting robust security measures.

## 7. Future Directions

This survey highlights significant vulnerabilities in Trusted Execution Environments (TEEs) stemming from physical and software-based fault injection (FI) techniques, such as Electromagnetic Fault Injection (EMFI). While progress has been made in identifying and mitigating some of these risks, further research is essential to address the evolving threat landscape and ensure robust security for TEEs.

- Advanced Countermeasure Development

    - Future research should focus on designing adaptive countermeasures capable of responding in real time to anomalies caused by FI attacks. This could include dynamic reconfiguration of TEE security protocols or automated isolation of affected components.

    - Exploring novel hardware designs that inherently resist fault injection, such as materials and structures with improved electromagnetic shielding properties, is critical for long-term security.

- Exploration of Emerging FI Techniques

    - As non-invasive methods, such as EMFI, continue to advance, future studies should aim to identify potential vulnerabilities that could arise from next-generation FI tools and techniques.

    - Investigating the implications of hybrid attacks that combine physical and software-based FI strategies will provide a more comprehensive understanding of TEE vulnerabilities.

- Enhanced Simulation Frameworks

    - Future frameworks should incorporate machine learning algorithms to predict fault injection parameters more efficiently and accurately, reducing the time and resources needed for experimentation.

    - Expanding simulation capabilities to cover a wider variety of TEE architectures and devices will allow for broader insights into vulnerabilities across different platforms.

- Standardisation and Policy Development

    - Future efforts should focus on establishing international standards for FI resilience, ensuring that manufacturers worldwide adopt consistent practices for TEE protection.

    - Collaborating with policymakers to create regulations that mandate robust fault injection testing and mitigation strategies for critical devices is essential for safeguarding national security.

- Interdisciplinary Research and Collaboration

    - Collaboration among experts in hardware engineering, software development, and cybersecurity policy will foster innovative solutions that address vulnerabilities from multiple angles.

    - Strengthening collaborations between academia, industry, and government will ensure that research outputs are translated into practical solutions that benefit society and industry.

- Ethical Considerations in FI Research

- Future research should prioritise developing ethical frameworks for the use of advanced FI tools, ensuring they are deployed responsibly to prevent misuse.
- Establishing best practices for responsibly disclosing vulnerabilities to manufacturers and stakeholders will promote trust and constructive progress in the field.

- Expanding TEE Applications
  - As TEEs become increasingly integral to IoT and smart device ecosystems, research should explore their specific vulnerabilities and mitigation strategies in these rapidly expanding domains.
  - Preparing TEEs to handle emerging challenges, such as quantum computing threats, will ensure long-term security and adaptability.

## 8. Conclusion

Trusted Execution Environments (TEEs) are indispensable for securing sensitive operations across modern computing platforms. However, as this survey has demonstrated, the growing sophistication of fault injection (FI) techniques, particularly non-invasive methods such as Electromagnetic Fault Injection (EMFI), presents significant risks to the confidentiality, integrity, and availability of TEE-protected data. These vulnerabilities expose critical gaps in the security of billions of devices, ranging from smartphones and IoT systems to cloud platforms and national infrastructure.

Through an in-depth analysis of physical and software-based FI techniques, this study has highlighted the need for more robust countermeasures, improved TEE architectures, and updated threat models that reflect the realities of emerging attack vectors. Current security approaches must evolve to incorporate adaptive defences, rigorous FI testing protocols, and enhanced simulation frameworks to mitigate these risks effectively.

The implications of these findings are far-reaching, impacting device manufacturers, policymakers, researchers, and critical infrastructure stakeholders. This survey emphasises the importance of interdisciplinary collaboration to develop holistic solutions that address vulnerabilities at both the hardware and software levels. By fostering partnerships between academia, industry, and government, the insights gained from this research can drive meaningful advancements in TEE security, ensuring resilience against current and future threats.

As TEEs continue to underpin secure operations across diverse domains, their protection is paramount for maintaining trust in the digital ecosystem. Addressing the vulnerabilities identified in this survey will not only safeguard critical data and operations but also reinforce the integrity of secure computing systems worldwide. By leveraging the recommendations and future directions outlined in this study, stakeholders can work together to build a more secure and resilient technological landscape.